\documentclass[10pt,letterpaper]{article}
\usepackage{opex3}
\usepackage{amsmath}
\newcommand{\ket}[1]{| #1 \rangle}
\newcommand{\bra}[1]{\langle #1 |}

\begin{document}




\title{Entangled Photon Polarimetry}

\author{Joseph B. Altepeter, Neal N. Oza, Milja Medi\'{c}, Evan R.
Jeffrey,\textsuperscript{\dag} \\and Prem Kumar}

\address{Center for Photonic Communication and Computing, EECS Department
\\ Northwestern University,  2145 Sheridan Road, Evanston, IL 60208-3118}
\address{\textrm{\textsuperscript{\dag}} Leiden Institue of Physics, Leiden University, Leiden,
Netherlands}

\email{joe.altepeter@gmail.com} 


\begin{abstract} We construct an entangled photon polarimeter capable of
monitoring a two-qubit quantum state in real time.  Using this polarimeter, we record
a nine frames-per-second video of a two-photon state's transition from separability
to entanglement.\end{abstract}



\section{Introduction}

Photonic entanglement is a fundamental resource for quantum information processing and
quantum communications \cite{mike_and_ike}.  Engineering suitable entanglement sources for a
particular application, or integrating those sources into a larger
system, however, can be a challenging experimental task.  Generating high-quality 
entanglement requires protecting against or compensating for decoherence, single-qubit
rotations, and partial projections.  
For both free-space \cite{ent0, ent1, ent2, ent3} and fiber/waveguide-based entanglement 
sources
\cite{kumar-entanglement, migdall, takesue, oband, cband}, this means 
compensating for any polarization rotations or decohering effects which may
occur in transit to a destined application.  In addition to the aforementioned static effects, it is necessary to
test the source's stability in the face of real-time system perturbations
such as
atmospheric turbulence or fiber breathing owing to environmental
fluctuations.  At present, the best
available technique for measuring two-qubit entangled states is quantum state tomography
\cite{tomo1, tomo2},
a procedure which can provide a precise reconstruction of the
quantum state, but which generally requires 5--30 minutes to
complete.  This long measurement time can make debugging systematic
experimental problems---particularly those with short timescales---challenging, if not impossible.

The field of classical optical communications has faced similar
problems when transmitting polarized light over long
distances.  A \emph{polarimeter} is a common tool which is used to debug unwanted
polarization rotations or depolarization effects
(i.e., polarization decoherence).  A polarimeter actively monitors the
polarization state of a classical optical field, providing an experimenter with
a real-time picture of the optical field's Stokes vector (i.e., its
polarization state).  Similarly, an \emph{entangled photon polarimeter}---a
measurement device capable of performing quantum tomographies and
displaying the reconstructed two-qubit states in 
real time---would be a valuable tool for optimizing and deploying entangled photon
sources.

In this paper we present the first experimental
implementation of an entangled photon polarimeter, which is
capable of displaying
nine reconstructed density matrices per second via complete quantum state
tomographies.  This represents a speed improvement of
2--3 orders of magnitude over the best quantum state tomography
systems currently in use in laboratories around the world.  Using this new tool, we
record the first live video---at 9 frames-per-second (fps)---of a two-photon quantum state's
transition from separability to entanglement.

\section{Two-Qubit Polarimetry}

Two-qubit polarimetry is a specific example of two-qubit
\emph{quantum state tomography}, a procedure for reconstructing an
unknown quantum state from a series of measurements (generally
either 9 or 36 coincidence measurements performed using two single-photon
detectors per qubit \cite{tomo2}), each performed on an
ensemble of identical copies of the unknown state.  Three key parameters
can be used to characterize any experimental apparatus for quantum
state tomography: the time required to complete the state reconstruction
procedure and
the accuracy and precision with which the reconstructed density matrix represents the
unknown quantum state.

\begin{figure}[t]
    \centering
         \includegraphics[width=\textwidth]{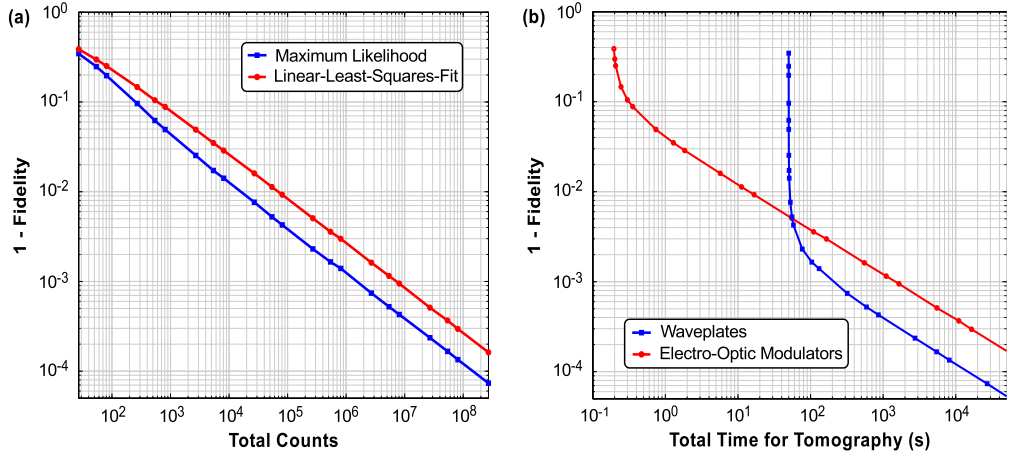}
         \vspace{-4mm}
    \caption{
        (a) 
        Tomographic precision $F_p ( N, \rho_\mathrm{ideal} )$ for 
        $\rho_\mathrm{ideal} = \ket{\phi^+}\bra{\phi^+}$ with 
        $\ket{\phi^+} = \frac{1}{\sqrt{2}} ( \ket{HH} + \ket{VV} )$.  Each
        data point represents a 2000-tomography numerical Monte Carlo
        simulation of the average fidelity between the reconstructed
        density matrix and $\rho_\mathrm{ideal}$, under realistic
        assumptions about the system noise (a coincidence-to-accidental ratio
        of 3). 
        Each simulated
        tomography utilizes four detectors and nine coincidence measurements,
        such that each unknown quantum state in the $N$-state ensemble is
        projected onto one of nine four-element orthonormal bases (e.g.,
        HH, HV, VH, VV).  Results for both the maximum likelihood technique
        and the truncated-eignevalue, linear-least-squares-fit
        technique are shown.  For a given $N$, the maximum likelihood
        technique is slightly more precise \cite{james-precision}.
        (b)
        Using the same simulated data, $F_p$ is shown as a function of total tomography time $T$ for two different
        experimental systems: a traditional free-space tomography system
        with $\eta = 0.1$, $\tau_s = 5$~s, $\tau_a = 5$~s and an
        entangled photon polarimeter with 
        $\eta = 0.07$, $\tau_s = 0.02$~s, $\tau_a = 0.001$~s.  In both
        systems $R = 10^6$ pairs/second and $M = 9$.
    }
    \label{figure::precision}
\end{figure}

The time required to complete a quantum state tomography,
$T$, is dependent on
the number of two-qubit measurement settings taken per reconstruction, $M$; the time
per measurement setting, $\tau_m$; the
time to switch between measurement settings, $\tau_s$; and the time
necessary to numerically reconstruct the unknown density matrix from an
analysis of the measurement results, $\tau_a$:
\begin{equation}
    T \equiv M \times (\tau_m + \tau_s) + \tau_a.
\end{equation}
The accuracy and precision of a tomography are closely related,
both indicating how closely the reconstructed density matrix, $\rho$, matches
the ``true'' unknown density matrix, $\rho_\mathrm{ideal}$.  The ``accuracy'' of a
tomographic reconstruction measures error due to systematic effects,
such as improperly performed projective measurements, uncharacterized drifts in
the detectors' efficiency, or a non-identical ensemble of unknown quantum states.
The ``precision'' of a tomographic reconstruction measures the
statistical error in $\rho$, and is strongly dependent on 
the total number of measurable states $N$ in the identical
ensemble (which is in turn dependent on the entanglement source's pair production rate,
$R$, and the total single-qubit measurement efficiency, $\eta$).  
In general, the tomographic precision 
decreases as $T$ (and therefore $N$) decreases \cite{james-precision}.  For sufficiently small
$T$ we can neglect systematic effects and quantify
tomographic \emph{precision} (as a function of $N$ and of $\rho_\mathrm{ideal}$) to
be the average fidelity between $\rho$ and $\rho_\mathrm{ideal}$:
\begin{equation}
    F_p \left( N, \rho_\mathrm{ideal} \right) \equiv
        \overline{F \left( \rho, \rho_\mathrm{ideal} \right)}
        = \overline{ \left( 
                \mathrm{Tr} \left\{
                    \sqrt{ \sqrt{\rho} \rho_\mathrm{ideal} \sqrt{\rho} } 
                \right\}
          \right)^2 }.
\end{equation}
Note that the equation above uses the usual definition for fidelity between two mixed
states \cite{fidelity}, which for a pure $\rho_\mathrm{ideal} \equiv
\ket{\psi}\bra{\psi}$,
simplifies to the more familiar 
$F \left( \rho, \rho_\mathrm{ideal} \right) \equiv \mathrm{Tr} 
    \left\{ \rho \rho_\mathrm{ideal} \right\} = \bra{\psi} \rho \ket{\psi}$.
Figure \ref{figure::precision}(a) shows $F_p(N)$ for 
$\rho_\mathrm{ideal} = \ket{\phi^+}\bra{\phi^+}$ 
with 
$\ket{\phi^+} = \frac{1}{\sqrt{2}} ( \ket{HH} + \ket{VV} )$, where each
data point represents a Monte Carlo simulation of the average fidelity
between a reconstructed density matrix and the ideal unknown state.

\emph{Two-qubit polarimetry} is an application of two-qubit
polarization tomography which maximizes precision for very short $T$ ($\le 1$s),
allowing an experimenter to manipulate an entangled photon source 
using real-time tomographic feedback 
(by updating after every measurement, the time between updates can be 
reduced to $T/9$).  
(In this paper, \emph{entangled photon polarimetery}
refers to the application of two-qubit polarimetry to entangled photon
states.)
Because maximizing precision
requires maximizing $N$, the ideal entangled photon polarimeter will
minimize both the time between measurements ($\tau_s$) and the time for
numerical analysis ($\tau_a$):
\begin{equation}
    N = R \eta^2 M \tau_m = R \eta^2 \left(
            T  
            - M\tau_s
            - \tau_a
        \right).
    \label{equation::N_T}
\end{equation}

Although Eq. (\ref{equation::N_T}) can be used to derive the total
time necessary to perform a single tomography with a given precision,
an entangled photon polarimeter will likely perform many
tomographies in series.  In this configuration, the tomographic
measurements and the numerical analysis of those measurements can be
parallelized in one of two ways.  For $\tau_a < M (\tau_m + \tau_s)$, a complete
set of $M$ measurements can be analyzed at the same time the next set of
$M$ measurements are being performed, leading to one tomographic result
being displayed
to the experimenter every $M ( \tau_m + \tau_s )$ seconds.  For $\tau_a <
\tau_m + \tau_s$, a tomographic result can be analyzed and displayed after
every \emph{measurement}, rather than after every complete set of $M$
measurements.  In other words, after every measurement, the \emph{previous
$M$ measurements} are used to reconstruct an updated density matrix, leading to a
faster refresh rate based on a tomographic ``rolling average''.  Similarly,
this configuration can be altered in real time to utilize even more
measurements (e.g., $4M$) for increased precision (analagous to
averaging multiple traces on an oscilloscope).


\section{Experimental Details}

The entangled photon polarimeter presented here is based on a previous
apparatus for free-space telecom-band quantum state tomography \cite{cband,
tomo2}, which
although accurate, is too slow to provide real-time feedback.  Three key
improvements have dramatically improved the tomographic speed while maintaining
precision: bulk wave plates have been replaced with fast
electro-optic modulators (EOMs), an array of four single-photon detectors
triggered at 8 MHz have been
replaced with an array that is triggered at 50 MHz, and the traditional
maximum likelihood reconstruction technique has been replaced
with a much-faster linear-least-squares-fit method.  

Below, we briefly discuss the differences between these two techniques
after reviewing the entangled photon source used to test the tomography
apparatuses.  Figure \ref{figure::precision}(b) highlights the differences
between the two techniques, showing the expected tomographic precision
$F_p$ as a function of total tomography time $T$.

\subsection{Entangled Photon Source}

\begin{figure}[h]
    \centering
         \includegraphics[width=\textwidth]{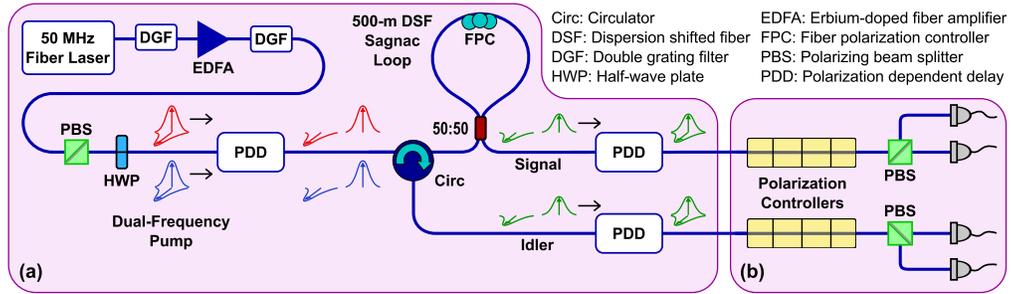}
    \caption{
    (a) The entangled photon source used to test the entangled photon
    polarimeter.  (b) The entangled photon polarimeter, composed of fast electro-optic modulators, in-fiber polarizers, and a four-detector array.
    }
    \label{figure::setup}
\end{figure}

To test the entangled photon polarimeter, we utilize a fiber-based,
frequency-degenerate, 1550-nm, polarization entangled photon-pair source
\cite{cband}.
The source utilizes spontaneous four-wave-mixing in dispersion-shifted
fiber and is pumped by 50-MHz repetition rate dual-frequency pulses
spectrally carved
from the output of a femtosecond pulsed laser. 
Because the output photons are identical, reverse Hong-Ou-Mandel
interference in a Sagnac loop is used to deterministically split the output
photons into separate output single-mode fibers.  See Fig.
\ref{figure::setup}(a).

The same source is used to test two separate tomography systems,
the automated wave-plate-based apparatus first described in
\cite{cband} and the entangled photon polarimeter presented here.  

\subsection{Polarization Measurements}

Traditionally, two-qubit polarization tomography is performed using bulk,
free-space, birefringent crystals (i.e., wave plates).  A quarter- and a
half-wave plate followed by a polarizer on each of the two qubits can implement an arbitrary
projective measurement \cite{tomo2}.  By collecting photons from both the
transmitted and the reflected
ports of each qubit's polarizer, one can project an unknown photon pair
into one of four orthonormal basis states, defined by the 
wave plates.  If well characterized, this can lead to a very accurate
tomography, though the measurement-to-measurement transition time 
$\tau_s$ will in general be very large ($\approx 5$s).  For the fiber-based
source above, this type of polarization analyzer will lead to a
single-qubit loss of $\approx 1.5$ dB (including the fiber to free-space to fiber
coupling losses).

To decrease $\tau_s$, we have constructed an all-fiber/waveguide polarization
analyzer based on electro-optic modulators (EOMs).  
These \(\text{LiNbO}_3\) EOMs (EOSpace, model PC-B4-00-SFU-SFU-UL) allow precise control of both the retardance and
optic axis of a birefringent crystalline waveguide using the fringe fields from three
electrodes.  In general, this process has an extremely short response time
leading to EOM switching rates of up to 10 MHz.  In practice, 
we are able to implement arbitrary
polarization measurements at 125 kHz, which is a limit set by the speed of our computer-controlled voltage
sources.  

Although high-speed, EOMs are more difficult to
precisely characterize than bulk wave plates; using a standard polarimeter
we have characterized the six transformations performed by each EOM-based
analyzer (corresponding to projections onto the H, V, 
$\textrm{D} \equiv (\textrm{H}+\textrm{V})/\sqrt{2}$, 
$\textrm{A} \equiv (\textrm{H}-\textrm{V})/\sqrt{2}$, 
$\textrm{R} \equiv (\textrm{H}+i\textrm{V})/\sqrt{2}$, and
$\textrm{L} \equiv (\textrm{H}-i\textrm{V})/\sqrt{2}$
basis states).  EOM projections deviated from an ideal
measurement by an average of 2.1 degrees on the Poincar\'{e} sphere.
The single-qubit losses of the EOM-based analyzers varied between
3.0--3.4 dB.

\subsection{Single-Photon Detection}

Single-photon detection is performed using a four-detector array of InGaAs
avalanche photodiodes (APDs) operated in the gated Geiger mode.  By increasing the speed of these detectors
from 8.3~MHz to 50~MHz, the entangled photon polarimeter achieves a 6-fold
speed increase relative to previous implementations of quantum state
tomography which utilized the same telecom-band detection systems.
Moreover, by 
synchronizing the detector-array's control software with the EOM-based analyzers,
we have reduced the switching time to $\tau_s = 20$ ms.  By upgrading the
detector control software to eliminate extraneous electronic delays, we
anticipate that this will approach the 125~kHz limit ($\tau_s = 10$~$\mu$s) imposed by the
EOM voltage controllers.
The quantum efficiency of each detector at 1550-nm is
approximately 20\%, with a measured dark-count rate of 1--4~$\times 10^{-4}$ per
pulse.

\subsection{Tomographic Reconstruction}

Traditionally, the
maximum likelihood technique has been used to reconstruct a two-photon
state's density matrix from a series of coincidence measurements, which
numerically solves for the density matrix
$\rho$ most likely to reproduce the measured counts \cite{tomo1, tomo2}.
This method always produces a legal state, but is relatively slow ($\tau_a \approx
5$~s).

By using a simpler analysis technique based on a linear least-squares
fit, we are able to increase the state reconstruction speed 
by more than three orders of magnitude \cite{james-precision}.  We use the 2-qubit Stokes vector
as a linear model, and solve
the following least-squares problem:
\begin{equation}
w M \cdot S = w C.
\end{equation}
Here, $M$ is the set of measurements, which can be arbitrary POVMs;  $C$ is
the measured counts, and $S$ is the Stokes vector we solve for;  $w$
is a weight vector representing the distribution width for each
measurement.  We assume the counting process to be Poissonian, and
use the large-N limit where the Poisson distribution is approximated
as a Gaussian with width $\sqrt{N}$.  To guarantee a legal density matrix,
we post-process the least-squares fit by truncating the negative eigenvalues
\cite{james-precision}.
We have found that this type of linear fit
provides results identical to those obtained via the maximum likelihood
method with a negligible drop in
precision (see Fig. \ref{figure::precision}(a)), only much
faster ($\sim 1.3$~ms per tomography using
Matlab on a 2.4-GHz CPU).

This three-order-of-magnitude speed increase allows us to display a new
frame (i.e., tomography result) after every measurement, reconstructed using the
previous $M$ measurements.  For four-detector, complete-basis
polarization analyzers (described above), only nine
measurements are needed to perform a complete tomography.
Note that it is often experimentally optimal to perform a redundant set
of 36 measurements in order to detect and/or correct for systematic errors
such as source intensity drift, detector efficiency drift, or polarizer
crosstalk \cite{tomo2}. 

\begin{figure}[t]
    \centering
         \includegraphics[width=0.9\textwidth]{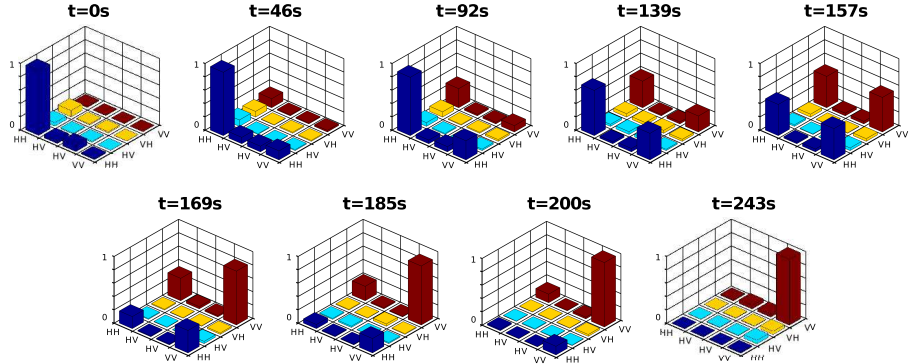}
    \caption{
        Selected frames from the nine fps video of a two-qubit photon
        state's transition from separability to entanglement.  
        Each frame shows a density matrix reconstructed
        using the previous 36 measurements ($\approx 4$~s of data).
    }
    \label{figure::movie}
\vspace{-5mm}
\end{figure}

\section{Entangled Photon Polarimeter Performance}

By utilizing fast EOM-based analyzers, a four-detector array triggered at
50 MHz, and a linear least-squares algorithm for tomographic reconstruction,
the entangled photon polarimeter is capable of performing nine tomographies
per second.  Operated at this speed, 
$\tau_m = 80$~ms, 
$\tau_s = 20$~ms, and
$\tau_a = 1$~ms.  Total single-qubit insertion loss is measured to be $\eta
= $~3--3.4~dB (not including detector inefficiency).  
The tomographic precision is estimated using a Monte Carlo 
simulation of this polarimeter's application to the entanglement 
source pictured in Fig. \ref{figure::setup} (resulting in
$\approx1000$ coincidences per second).
For nine-measurement tomographies ($T \approx 1$~s), $F_p ( \rho,
\rho_\mathrm{ideal} ) \approx 92\%$. 
For 36-measurement tomographies ($T \approx 4$~s), $F_p (\rho,
\rho_\mathrm{ideal} ) \approx 96\%$. 

To experimentally verify this performance, we recorded three 9-fps live
videos of a two-photon polarization state using the 36-measurement
configuration.  First we recorded two videos where the measured state is
not changed during the course of the measurement run, for a totally separable pure state,
$\ket{HH}$, and a maximally entangled state,
$\ket{\phi^+}$.  By analyzing each frame and
comparing it to the target state, we directly measured the system
precision to be $98\% \pm 1\%$ (for $\ket{HH}$) and $95\% \pm 2\%$ (for $\ket{\phi^+}$).
Finally, we recorded a video of a two-photon state's transition from
separability to entanglement (the transition is physically implemented by
rotating the wave plate HWP in the entangled photon source setup---see Fig.
\ref{figure::setup}).  Selected
frames from this video are shown in Fig. \ref{figure::movie}.

This research was supported in part by the DARPA ZOE program (Grant No.
W31P4Q-09-1-0014) and the NSF IGERT Fellowship (Grant No. DGE-0801685).

\end{document}